\documentstyle[11pt,aaspp4]{article}
\input epsf
\lefthead{SANDQUIST ET AL.}
\righthead{DOUBLE CORE EVOLUTION. XI. }
\def\n{\noindent}
\def\msun{\mbox{~M$_{\odot}$}}
\def\rsun{\mbox{~R$_{\odot}$}}

\def\lapprox{\;\rlap{\lower 3.0pt                       
             \hbox{$\sim$}}\raise 2.5pt\hbox{$<$}\;}
\def\gapprox{\;\rlap{\lower 3.0pt                       
             \hbox{$\sim$}}\raise 2.5pt\hbox{$>$}\;}

\def\ace{$\alpha_{CE}$}
\begin{document}
\title{ON THE FORMATION OF HELIUM DOUBLE
DEGENERATE STARS AND PRE-CATACLYSMIC VARIABLES}
\author{Eric L. Sandquist\altaffilmark{1,2}, Ronald E. Taam\altaffilmark{1},
and Andreas Burkert\altaffilmark{3}}

\n \altaffilmark{1}{Department of Physics \& Astronomy, Northwestern
University, Evanston, IL 60208; erics@mintaka.sdsu.edu,
taam@apollo.astro.nwu.edu}

\n \altaffilmark{2}{present address: Department of Astronomy, San Diego State 
University, 5500 Campanile Drive, San Diego, CA 92182}

\n \altaffilmark{3}{Max Planck Institut for Astronomy, Koenigstuhl 17, 
D-69117 Heidelberg, Germany; burkert@mpia-hd.mpg.de}

\begin{abstract}

The evolution of low mass (M $< 2.5 \msun$) binaries through the
common envelope phase has been studied for systems in which one member
is on its first ascent of the red giant branch. Three-dimensional
hydrodynamical simulations have been carried out for a range of red
giant masses ($1 - 2 \msun$) with degenerate helium cores ($0.28 -
0.45 \msun$) and companions ($0.1 - 0.45 \msun$) for initial orbital
periods ranging from $\sim 15 - 1000$ days.  The results suggest that
these low mass binary systems can survive the common envelope phase
provided that the helium degenerate core is more massive than about
$0.2 - 0.25 \msun$ and that the mass of the red giant progenitor is
$\lapprox 2 \msun$.

Specific applications are made to observed double helium degenerate
systems, pre-cataclysmic variables, and subdwarf B stars in order to
place constraints on progenitor systems evolving through the common
envelope phase. For the observed short period double degenerate
systems, it is found that evolutionary scenarios involving two phases
of common envelope evolution are not likely and that a scenario
involving an Algol-like phase of mass transfer followed by a common
envelope phase is viable, suggesting that the first-formed white dwarf
is often reheated by nuclear burning on its surface.  A formation mechanism
for two subdwarf B stars observed in eclipsing short period
binaries with low mass main sequence stars is also described.
\end{abstract}


\section{Introduction}\label{intro}

Short-period ($P \sim$~hours) binary systems involving low-mass white
dwarf stars have been discovered in increasing numbers over the past
five years (e.g. Marsh, Dhillon, \& Duck 1995; Marsh 1995). Because
white dwarfs can only be created at the cores of giant stars, this
indicates that the orbital separation of the components must have been
much greater in the past. This situation is generally thought to be
the result of common envelope (CE) evolution (Paczynski 1976), in
which unstable mass transfer from a giant star to a companion causes
the companion to be engulfed. The ejection of the giant's envelope
requires energy to be removed from the orbit, resulting in a
short-period binary involving a white dwarf --- what had
been the core of the giant. For those white dwarfs with mass $M
\lapprox 0.5 \msun$, the giant star is required to be on its first ascent
up the red giant branch (Iben, Tutukov, \& Yungelson 1997).

The efficiency at which the gas is ejected from the giant's envelope
(\ace) is a critical parameter for determining the final orbital
period of such systems. We use the definition \[ \alpha_{CE} = \frac{\Delta
E_{bind}}{\Delta E_{orb}}, \] where $\Delta E_{orb}$ is the change in
orbital energy of the binary [from (giant, companion) 
to (white dwarf, companion)], and
$\Delta E_{bind}$ is the binding energy (as determined from the
initial giant model) of mass unbound during the simulation.
Because the evolution is rapid (making radiative transfer
effects unimportant during the main mass ejection phase; Sandquist et
al. 1998), the evolution is adiabatic.  The only inefficiencies that
affect calculations of \ace ~are kinetic energy given to gas during
the hydrodynamical interaction beyond what is necessary to unbind it,
and kinetic energy given to center of mass motion of the remnant
binary.

To determine \ace ~for a real system, the final masses of the two
components of the binary, the final orbital period, and the initial
mass of the giant's envelope should be known at least.  The progenitor
binary can be described by the initial mass of the giant $M_{g}$, its
core mass $M_{c}$, its radius $R$, and the mass of the companion
$M_{2}$.  The initial orbital separation $a_{i}$ need not be specified
since it can be obtained assuming the giant fills its Roche lobe.  The
giant's radius can be estimated using a core mass-radius relation for
a red giant if one component of the binary can be identified as the
giant's core. The initial mass of the giant envelope $M_{env}$ is the
greatest uncertainty since there are no obvious constraints except
from measurement of nebular mass. However, a lower limit on the total
mass of the giant can be obtained by requiring that it evolved within
the age 
of the Galactic disk (although the giant mass could be even lower if
there is an effective stellar wind prior to the common envelope
event). So, at the very least we require observed binary systems with
well-measured orbital periods and component masses in order to
estimate \ace.

We have attempted to model some of the best studied low-mass binary
systems that meet these criteria.
In this paper, we consider two types of systems containing helium
white dwarfs. The first is the pre-cataclysmic variable stars, which
are detached binary stars involving a white dwarf and a low-mass main
sequence star. Subsequent angular momentum loss
due to magnetic braking and gravitational radiation can reduce the
orbital separation until the main sequence star exceeds its Roche
radius and initiates mass transfer. The second class is double
degenerate systems, in which both components are helium white
dwarfs. The period of such systems determines whether the two
white dwarfs can merge in less than a Hubble time through
gravitational radiation. Because of the low white dwarf masses, the
merger of components will not typically form a Chandrasekhar mass
object and supernova. Although such objects may lead to helium detonation, such
systems can compose at most about 10\% of observed Type I supernovae
(Woosley, Taam, \& Weaver 1986). This percentage follows from the
overproduction of $^{44}$Ca relative to iron (assuming that Type I supernovae
account for the Galactic iron abundance), and is consistent with estimates by
Iben \& Tutukov (1984). The orbital parameters of the
systems we have considered are given in Table~\ref{obs}. 

In \S 2, we briefly
describe our computational method and initial models. In \S 3, we
discuss the general features of the hydrodynamical evolution of the
system, and highlight features of individual
simulations. In \S 4, we discuss observed binaries containing
helium white dwarfs, and the constraints our
models place on the process of common envelope evolution and 
on formation mechanisms for real binary systems.

\bigskip

\section{Numerical Methods}

The numerical techniques we have used in the simulations presented
here are largely identical to those used in Sandquist et al. (1998)
and discussed by Burkert \& Bodenheimer (1993). We briefly summarize
the method below, and enumerate the improvements made since the
previous study.

The computational regime was composed of a 3-D grid of $64 \times 64
\times 64$ zones, with two subgrids of $64 \times 64 \times 32$ zones
nested within, and centered on the main grid.
The subgrids were a factor of 2 smaller in the $x$ and $y$ directions
compared to the next coarser grid. Two point masses were used to
represent the degenerate core of the red giant star and its companion
(which can be identified as either a main-sequence star or helium
white dwarf).  Gravitational interactions between the collisionless
particles and the gas were smoothed according to \[ \Phi_{PG} =
\frac{-G M_{P}}{\sqrt{r^{2} + \epsilon^{2} \delta^{2}
\exp(-(r/\epsilon \delta)^2)}} ,\] where $\delta$ is the width of a
zone on the innermost subgrid, $\epsilon = 2.0$, and $r$ is the
distance between the gas and the particle of mass $M_P$.  The
gravitational potential of the gas was computed via fast Fourier
transform solution of Poisson's equation.

The orbit of the core particles decayed under the action of
gravitational drag forces within the common envelope. The core positions
were obtained using a four-point Runge-Kutta integration scheme each
time the gas distribution (and thus, the gravitational force) was
recalculated on the innermost subgrid. When the cores were at their
closest approach, the gas timestep was subdivided to ensure that the
core positions were computed at least 25 times during each orbit.

\bigskip

\subsection{Initial Models}

The initial gas distribution in the envelope of the red giant was
taken from one-dimensional stellar models obtained from the code
developed by Eggleton (1971, 1972) and updated by Pols et
al. (1995). In order to stabilize the giant model once the core
particle had been inserted, we adjusted the temperature of the 8
subgrid zones immediately surrounding the core to put them into
hydrostatic equilibrium. The remainder of the grid was filled with a
diffuse background gas in pressure equilibrium with the surface of the
star.

The companion was typically placed on a circular orbit in the
$xy$-plane with a radius about 30\% larger than the surface of the
giant as a compromise between computational time and matching the
initial conditions for a real binary beginning mass transfer. The
difference is unimportant because the deep structure of the giant
(where most of the energy input occurs in our simulations) is
unchanged by the difference in initial separation. The envelope of the
giant star was given a rotation rate that would have made it
synchronous with the companion if the companion was on an orbit that
would cause the giant to just overflow its Roche lobe. As a result,
the envelope was out of synchronicity with the companion. This was
done because tidal forces are not able to maintain synchronism between
the giant's envelope and the companion once mass transfer begins
accelerating.

\section{Results}

The initial parameters and results for each sequence are summarized in
Tables~\ref{initialt} and \ref{finalt} respectively.  For all of the
simulations that we have carried out in this study, the initial phases of the
interaction between the giant star and its companion are similar to
what was reported in Sandquist et al. (1998).

\subsection{The Spiral-In and Rapid Infall Phases}

Since these phases were discussed in more detail in Sandquist et
al. (1998), we only summarize the general features here.  In the
earliest portion of the evolution, gas is pulled from the surface of
the giant to the companion and is gravitationally torqued, removing
angular momentum from the binary orbit until the companion and giant
come into direct contact. Once this occurs, the orbital decay
accelerates for a period of time.  Large-scale spiral shock waves
appear, but soon become tightly wound as the period of the binary
orbit becomes smaller than that of the envelope gas. Once the two
point masses begin to interact with higher density gas, the rate of
energy transfer to the gas increases.

There are two distinct intervals during the evolution when gas is
unbound in the system.  The first occurs as the companion slingshots
gas into a spiral wave during its descent through the giant
envelope. For a relatively long period after the initial infall, the
amount of bound mass remains nearly constant, although there is still
some expansion of the gas (as can be seen in Figures~\ref{mass2}, \ref{mass3}, and \ref{mass4} in
the decrease in the amount of mass retained within the original volume
of the star, $M_{vol}$). More substantial ejection of mass
begins when the point masses are deep in the potential
well of the binary, creating a more vigorous outflow in the
orbital plane.

\subsection{The Envelope Ejection Phase \label{ee}}

The final separation of the binary is the most important quantity that
can be derived from simulations. In our previous study, as well as for
most of the sequences presented here, the resolution of the gas in the
grid limited how long we could accurately compute interactions between
point masses and the gas. For this reason most of the simulations had
to be stopped at or before the start of the envelope ejection phase
(defined by Sandquist et al. 1998 as the time at which the orbital
decay timescale begins rising after the rapid infall). In sequence 3,
however, this was not a factor until a much later stage in the
computation --- primarily because the binary had expelled most of the
giant's envelope before this point.

Previous papers in this series have indicated that the initial
structure of the red giant plays an important role in the orbital
decay. In an evolved giant branch star, there is an extremely small
amount of mass in the radiative region between the hydrogen-burning
shell and the base of the convection zone, even though that
region encompasses a large volume. When the orbital
separation becomes comparable to the size of this low mass region, the
orbital decay will slow.

Simulation 3 shows that the orbital decay can be decelerated in a
completely different way. From one-dimensional models of the initial
giant, the base of the convection zone is $1.5 \times 10^{12}$~cm from
the core. However, when the orbital separation reaches this size, the
orbital decay timescale has increased to 4500 days, indicating that
the binary is essentially stable (although the giant's envelope has
not been unbound and ejected from the grid). This appears to show that
with a relatively high-mass binary (compared to the mass of the
giant's gas envelope) gravitational torques can remove the gas and
halt the orbital decay before it reaches what was the radiative zone
of the giant.

In order to put all of our simulations in a theoretical context, we
have computed the value of the ratio $$\gamma_{CE} \equiv
\frac{\tau_{spin-up}}{\tau_{decay}}$$ (Livio \& Soker 1988) from the
stellar models we used to initialize the hydrodynamics.  $\gamma_{CE}$
is a measure of the importance of the angular momentum transfer to the
giant envelope. For $\gamma_{CE} < 1$, significant spin-up of the
giant's envelope is likely during the decay of the companion's
orbit. Assuming that the giant is not rotating, and that the companion
is moving on a circular orbit at its Keplerian speed, the ratio
becomes \[ \gamma_{CE} \simeq \frac{4}{5} \left(
\frac{\tilde{\rho_{a}}}{\bar{\rho_{a}}} \right) \left( \frac{M_{1}(a)
+ M_{2}}{M_{2}} \right) \] where \[ \tilde{\rho_{a}} = \frac{5}{a^{5}}
\int^{a}_{R_{core}} r^{4} \rho (r) dr, \] $R_{core}$ is the radius of
the core of the giant, $M_{1}(a)$ is the mass of the giant's envelope
within the orbit of the companion, and $\bar{\rho_{a}}$ is the average
density of mass inside the orbit. We are in a position to compare the
predictions of this simple model to the results of our hydrodynamical
simulations.

Figure~\ref{gamma} shows plots of $\gamma_{CE}$ as a function of
radius in the envelope of the giant for each of our simulations. To
check the validity of the simple spin-up model, in Figure~\ref{rhoend}
we plot slices for all of our simulations at the end of the rapid
infall phase. The assumption of a Keplerian orbit for the companion is
clearly violated at the times depicted, but the majority of the
angular momentum transfer has occurred earlier. The simple model only
predicts significant spin-up for simulations 2, 3, and 6, with
simulation 3 being the most likely to show the effects. The more
massive the companion and the lower the moment of inertia (whether due
to low envelope mass or relatively small radius), the more likely
spin-up will occur. In agreement with the prediction, simulation 3
shows dramatic effects of spin-up of the gas (including depletion of
gas along the binary axis, and stable rotation), while simulations 2
and 6 show these features to a lesser extent.

\bigskip

\subsection{Discussion of Individual Simulations}

In Table~\ref{finalt}, we list the separation and period of the point
mass binaries at the end of our simulations. We emphasize that for all
but simulation 3 the binary would undergo additional orbital
decay. For that reason, the values listed for the efficiency of mass
ejection \ace~ should only be considered estimates. Our
resolution prevents us from running our simulations far enough into
the envelope ejection phase to accurately determine \ace due to
either rapid orbital decay or to slowly evolving gas on the grid.

Instead we will focus our discussion on whether the simulated binary
is likely to survive or merge during the common envelope phase, and
what conclusions can be drawn for real binary systems involving helium
white dwarfs. Although we cannot make good estimates of \ace, we can
make a crude estimate of the final orbital separation for the binary
using the energy needed to unbind the envelope mass remaining
at the end of our simulations. Assuming values of \ace ~$= 0.4$
(Sandquist et al. 1998) and 1.0, we can compute how far the orbit
would have to shrink to release the energy needed to
complete the ejection.

\subsubsection{Simulation 1: $2 \msun$ Giant, $0.2 \msun$ Companion}

This simulation was run primarily to examine a case in which the
binary was likely to merge because the low-mass remnant binary has
insufficient energy to eject the relatively large envelope mass.
Figure~\ref{times1} compares the orbital decay timescale for the
binary with the timescale for change in bound mass.  The orbital decay
timescale at the end of the simulation was several thousand times
shorter than the timescale for unbinding and ejecting mass from the
system, indicating that the binary is likely to merge.

As seen in Figure~\ref{mass1}, very little mass had been unbound by
the end of the simulation. The amount of energy necessary to unbind
the remaining gas was over 10 times greater than what was removed
during the simulation. Our estimates of the final orbital separation are
also quite low --- only a few tenths of a solar radius. We conclude that this
system is indeed likely to merge.

\subsubsection{Simulation 2: $1 \msun$ Giant, $0.35 \msun$ Companion}

This binary was modeled as a plausible progenitor of the double
degenerate system PG 1101+364, which has $q=0.87\pm0.03$, a period of
3.47 h (Marsh 1995), and a mean mass of approximately $0.31 \msun$
(Bergeron et al. 1992). The temperatures of the two components of the
binary leave some confusion as to the formation mechanism for this
system (see \S~\ref{dd}), but its short period indicates that it
probably evolved through a common envelope phase. WD 0957-666 may be a
better system for comparison --- it also has a short period (1.46 h)
and similar masses ($0.37\pm0.02 \msun$ and $0.32\pm0.03 \msun$; Moran
et al. 1997).

A modest amount of energy is necessary to completely eject the
envelope mass remaining at the end of this simulation, so the
estimated final orbital separation for the system is not much
different from the separation that was reached. Based on this, it is
likely that this system will survive, particularly if both of the
components are helium degenerates. Figure~\ref{mass2} corroborates
this, showing that most of the gas has been removed from the original
volume of the star, including the dense gas around the giant's
core. The estimated final periods for this simulation compare well
with the observed double degenerate systems.

\subsubsection{Simulation 3: Evolved $1 \msun$ Giant, $0.35 \msun$ Companion}

This simulation was intended for comparison with the pre-CV system
PG 1224+309, which has an orbital period of 6.21 h, a white dwarf with
a mass of approximately 0.45 \msun, and an M dwarf companion of
$0.28\pm0.05 \msun$ (Orosz et al. 1999). At present, PG 1124+309 is the
pre-CV system with main sequence companion of largest measured mass,
although CV companion masses can be somewhat larger.  The mass of the
giant is approximately the minimum for a star that could have evolved
to the giant branch in the age of the Galactic disk. 
Together with the highly
evolved state of the giant (low mass and binding energy of the giant
envelope) and the relatively massive remnant binary, this
represents a system that is likely to survive the common envelope phase.

Because the orbital decay of the binary slowed substantially before
reaching the point where grid resolution became important, we
were able to follow this simulation much farther in its evolution than
any other.  Several features distinguish this simulation from the
previous ones.  As can be seen from a cut through the $yz$-plane
in Figure~\ref{rhoend} (panel c), there is a very low
density region oriented along an axis perpendicular to the orbital
plane of the binary. In addition, we find that the circulation pattern
that is often seen in common envelope simulations (e.g. Sandquist et
al. 1998) is almost completely absent in this simulation because of
the support given to the gas by centrifugal forces. During a portion
of the simulation, a large and stable differentially rotating
disk-like structure occupies much of the computational domain.

As can be seen from a plot of mass tracers in Figure~\ref{mass3},
negligible mass is unbound after the initial spiral-in
phase. Towards the end of our simulation, the mass distribution
appears to stabilize for a time. However, beginning at approximately
900 d, convective plumes appear and begin to disrupt the
disk. Stability of the material against local axisymmetric adiabatic
perturbations was tested using the H{\o}iland criterion (Tassoul 1978; Igumenshchev,
Chen, \& Abramowicz 1996):
\[ \frac{1}{r^{3}} \frac{\partial l^{2}}{\partial r} - 
\left(\frac{\partial T}{\partial P}\right)_{S} \nabla P \cdot \nabla s > 0\]
and
\[ -\frac{1}{\rho} \frac{\partial P}{\partial z} 
\left(\frac{\partial l^{2}}{\partial r} \frac{\partial s}{\partial z} -
\frac{\partial l^{2}}{\partial z} \frac{\partial s}{\partial r}\right) > 0\]
where $l$ is the specific angular momentum, and $s$ is the specific
entropy. By the H{\o}iland criterion, the plumes we see in our
simulation are indeed unstable. The unstable regions for one
timestep are plotted in Figure~\ref{hoil}.

At the end of the simulation the total bound mass (the sum of what
remained on the grid and what moved off the grid) had decreased only by
about $0.05 \msun$ down to $0.5 \msun$. However, the gas contained within the
original volume of the star had decreased to about $0.1 \msun$, so that
most of the gas had expanded to relatively large
distances from the point masses. The final orbital period of the point
masses was 13 days, indicating that it may
be possible to create binaries with periods of several days in the
common envelope scenario. The orbital decay timescale of the binary
had also risen to about 4500 days. Given the low binding energy of the
gas remaining in the simulation, the binary is likely to remain with
an orbital period on the order of several days. A period on the order of days
is considerably larger than that of PG
1224+309, which suggests that the mass of the giant's envelope
must have been larger.  This possibility is examined in simulation
5. However, to make a definitive statement about the final state of
the binary, we need to understand the role of the gas that
remained on the grid. This gas influences the
evolution on timescales longer than we are able to follow.

Although this simulation is best suited for a useful computation of
\ace~, the calculation yields a rather low value (0.09).  In this
case, the efficiency is low because the gas has been moved
far from the cores without being
unbound. This is a consequence of our definition of \ace, and is not
necessarily a reflection of the efficiency of mass ejection for 
common envelope systems.

\subsubsection{Simulation 4: Evolved $1 \msun$ Giant, $0.1 \msun$ Companion}

This run is to be compared with the pre-CV system
GD 448, which has an orbital period of 2.47 h, a white dwarf with a
mass of approximately $0.44\pm0.03 \msun$, and an M dwarf companion of
mass $0.09\pm0.01 \msun$ ~(Maxted et al. 1998). It may also
simulate the formation of the subdwarf
B/main sequence star binaries HW Vir (Wood, Zhang, \& Robinson 1993)
and PG 1336--018 (Kilkenny et al. 1998), which have periods of 2.8 and
2.4 hours respectively. The sdB/MS binaries will be discussed in more
detail in \S~\ref{sdb}, but sdB stars are generally believed to have a
mass of about $0.5 \msun$ (Saffer et al. 1994), and mass ratios $q
\sim 0.3$ are inferred in the cited observational studies. The companion mass
and initial orbital separation are the only differences between the
initial parameters of this simulation and simulation 3. The companion
mass is near the hydrogen burning limit for stars, and is at the low end of the
companion mass distribution for well-observed pre-CV systems.

The orbital decay timescale for the binary at the end of the
simulation is about 200 days, while the mass unbinding timescale is
considerably longer (and is not well-determined due to the low rate of
change of the bound mass). The mass tracers for this simulation are
shown in Figure~\ref{mass4}. Very little mass ($\sim 0.02 \msun$) was
made unbound by the time the simulation was terminated, and the
majority of the gas originally within the volume of the giant
remained there. As a result, the binary is certain to undergo
substantial further orbital decay. 

The estimated final orbital periods range from 0.8 to 2.6 days, which
are again substantially larger than those of the observed systems. A
common envelope binary involving a higher mass giant in the same
evolutionary phase could potentially produce the observed
systems. However, we must be careful: if the \ace ~values we have used
are overestimates, the final periods would also be smaller. Because
this simulation shows less evidence of mass ejection than simulations
2 and 3, and because the low mass of the companion makes it more
likely that the binary will merge, it is not certain that
\ace ~will be in the range we used to calculate the periods. Better resolved
computations should be undertaken.

\subsubsection{Simulation 5: Evolved $2 \msun$ Giant, $0.35 \msun$ Companion}

In order to examine the
effects of giant mass on the binary, we reran simulation 3,
doubling the giant mass while leaving the mass of the giant core and
companion the same. This places the giant mass near the maximum that
can produce a helium degenerate object.

The additional gas mass drastically changes the outcome.  The
timescale for orbital decay at the end was approximately 100 days,
while the mass-loss timescale was around $10^4$ days. Energetic
considerations indicate that the binary is likely to shrink to a
separation of a few solar radii. However, the caveats from simulation 4 can be
applied here also --- unless the value of \ace~ is proven to be
between 0.4 and 1.0 (what we consider to be the most likely range), we
cannot be certain that the mass ejection is less efficient than in
previous simulations. As a result, we believe that the estimates of the
final orbital separation in Table~\ref{finalt} are upper
limits. Those upper limits are still greater than is found for the
observed systems, so it is still possible that a giant of this
mass could produce them.

\subsubsection{Simulation 6: $1 \msun$ Giant, $0.45 \msun$ Companion}

This simulation was run as one test of the validity of the double CE
mechanism for forming double helium white dwarf systems. It is
intended to model a binary after it has undergone the first common
envelope phase, which shrank the orbit by a large factor. As a result,
the secondary would not have been able to evolve as far before mass
transfer began. Here we assume the giant has been able to form a
$0.28 \msun$ core, putting it on the lower half of the red giant
branch.

This binary is likely to survive the common envelope interaction ---
by the end of the simulation only $0.25 \msun$ of the original $0.72
\msun$ of gas was left within the original volume of the giant. Because
of the evidence that the system will eject much of its mass, we can be
more confident in believing the range of final periods in Table~3.
These estimates of the final period fall on
the order of a few hours, which is comparable to the short period
double degenerates in Table~1.

\section{Discussion}

\subsection{Constraints on CE Evolution}

The common envelope interaction of binary stars with a low-mass (M $<
2.5 \msun$) red giant has been investigated for the formation of
double degenerate systems and pre-cataclysmic variable systems.  The
results of three-dimensional hydrodynamical simulations suggest that
common envelope evolution is a viable method of producing these
systems. 
For red giants with helium white dwarf cores
$M_{c} \lapprox 0.2 - 0.25 \msun$, the absence of an extensive region
characterized by a flat mass-radius profile inhibits the survival of
the system as a binary (see Figure~\ref{mrds}). 
Although an extensive set of calculations has not yet demonstrated that 
all red giants of low core mass merge with their companions, it is likely
in this case that gravitational torques remain effective in bringing the 
two cores together, which may 
lead to the dissolution of the companion and to the
formation of a rapidly rotating subgiant star.  
If so, then the two stars would continue
to spiral together {\it even} if sufficient energy is available to
unbind the envelope since the timescale for orbital decay is 
expected to be 
shorter than the timescale for mass loss from the common envelope.
In this interpretation, the observational detection of a system 
containing a white
dwarf less massive than $\sim 0.2 - 0.25 \msun$ in double degenerate
and pre-cataclysmic systems would imply the existence of an
alternative evolutionary path.  Such a path might involve substantial mass
and angular momentum loss via an Algol-type evolution for the double
degenerate systems (Sarna, Marks, \& Smith 1996).

When the red giant star is in an advanced evolutionary state, a larger
fraction of the star's mass is stored at larger radii. As shown in
this paper (simulation 3, for example), the envelope can be pushed 
farther outward by the interaction between the envelope and the two cores,
provided that the cores are sufficiently massive.  In that case,
gravitational torques are less effective in forcing the two
stars together, and energy considerations become the important
determinant of the survival of the binary. Our simulations
indicate that for a system to survive as a double degenerate,
the giant's envelope mass at a given evolutionary stage is
constrained by the mass of the companion white dwarf.

The energy release associated with hydrogen recombination has
previously been suggested as an additional source of energy for
ejection of the common envelope. For planetary
nebulae, detailed calculations have failed to demonstrate that
hydrogen recombination can cause mass ejection (see Harpaz 1998).  
The reduction in
opacity associated with the recombination increases radiative energy
losses, thereby reducing the efficiency of the mass ejection
process. Although recombination may play a minor role in helping eject
the outer surface layers of the giant, it is likely to be ineffective
in unbinding matter at higher temperatures in the deep interior of the
common envelope.

\subsection{Constraints on Formation Mechanisms}

The success in ejecting the common envelope during the red giant stage
provides theoretical support to the hypothesis that the common
envelope phase can produce low-mass double degenerate and
pre-cataclysmic systems.  The primary difference between double
degenerates and the pre-cataclysmics is the evolutionary state of the
companion --- for pre-CVs, the companion is a main sequence star,
which is considerably larger than a white dwarf. Thus, for pre-CVs,
orbital separations must be such that the tidal lobe is not smaller
than the thermal equilibrium radius of the main sequence star.

\subsubsection{Pre-Cataclysmic Variable Stars}

For pre-cataclysmic variable systems containing a white dwarf and a
detached low-mass main sequence companion, we are assured that only
one CE phase can have occurred. There are two well-studied systems of
this type that we can compare with our simulations. Maxted et
al. (1998) observed GD448 (WD 0710+741), a system containing a white
dwarf and an M dwarf companion orbiting with a period of 2.47
h. Although the period places it in the CV period gap, making it
possible that GD 448 is in a temporary detached state, Marsh \& Duck
(1996) argue that it was born into this state following a common
envelope phase. Their arguments are based on the short cooling time
for the white dwarf ($5 \times 10^{7}$~yr) compared to the timescale
for angular momentum loss via gravitational radiation ($1.8 \times
10^{9}$~yr) necessary to reduce the period to the present value from
the 3 h upper edge of the period gap. (Note that some of the orbital
evolution may have taken place via magnetic braking.) Orosz et al.
(1999) studied the system PG 1224+309, which orbits every 6.21 h.

The masses of the white dwarfs in both systems indicate that the
giants were approaching helium flash when they formed a common
envelope.  This fact ($M_{c} \gapprox 0.4 \msun$) implies that the
giant's envelope had relatively low binding energy (radius of $\sim
10^{13}$ cm), enabling a low-mass main sequence companion to eject the
envelope.  In order for the giant to have evolved to this state, its
main-sequence evolutionary timescale must be less than the age of the
Galactic disk ($\approx 10$~Gyr), which provides the constraint that
$M \gapprox 1.0 \msun$. Our numerical results indicate that a binary
containing a $0.35 \msun$ companion and an evolved $1 \msun$ giant at
a period of several days is highly likely to survive.  The fact that
the final orbital period found in simulation 3 is considerably longer
than that of PG 1224+309 indicates either that the mass of the giant
star was larger than $1 \msun$, or the gas remaining on the grid will
interact with the binary to further reduce the orbital separation. Our
knowledge of the mass of the white dwarf in this system fixes the
evolutionary state of the giant.

With regards to the formation of pre-CVs from an initial binary
composed of a $1 \msun$ giant (with a $0.45 \msun$ core) and a $0.1
\msun$ companion, our results are uncertain.  The existence of GD 448
indicates that such systems probably do survive.  It is possible
though that mass loss on the giant branch (e.g., Reimers 1975, 1977)
may have reduced the envelope mass prior to the onset of the common
envelope phase. The energy required to eject the envelope would be
reduced, and survival of the binary could become possible.  The
importance of mass loss for single red giant branch stars has long
been realized because it appears to be needed to explain the masses of
horizontal branch stars in globular clusters (Rood 1973). Other
evidence provided by the width of the color distributions in these
clusters suggests that there are star-to-star differences in the
amount of mass loss (Rood 1973). In addition, a metallicity dependence
might be expected (Renzini 1981), which would imply that mass loss
among stars in the Galactic plane is likely to be larger than for
stars in globular clusters on average. The existence of a companion in
close proximity to the giant could further enhance the mass loss (Tout
\& Eggleton 1988) and facilitate the successful ejection of the common
envelope.

\subsubsection{Double Degenerate Systems\label{dd}}

In the following, we discuss three possible mechanisms for forming
binary helium white dwarfs, which we shall refer to as the Algol/CE,
CE/CE, and CE/Algol mechanisms.

In the Algol/CE scenario, the primary begins mass transfer as a
subgiant (with an initial separation $a_{i} \lapprox 20
\rsun$). Provided that the system does not undergo a dynamical
instability (which can occur if the mass ratio exceeds a critical
value that depends on the adiabatic response of the star), further
evolution results in mass transfer on a nuclear timescale as an
Algol-like system. The donor's envelope is eventually depleted to the
point that it contracts within its Roche lobe and is seen as a white
dwarf. By the end of this stage, the orbital separation has increased
beyond its initial value. The system remains detached until the
secondary evolves onto the first-ascent giant branch and fills its
Roche lobe.
If the convective envelope of the (now more massive) donor is sufficiently
massive, mass
transfer is unstable and the system enters into the common envelope
phase.  Envelope ejection takes place, and the final binary separation
is determined by the evolutionary state of the giant and the amount of
energy released from the binary orbit.

The mass ratio of a system can serve as an important clue
to the formation channel. For the Algol/CE scenario, the mass ratio
of the fainter to the brighter white dwarf is constrained very tightly
to $q=0.88\pm0.03$ (Tutukov \& Yungelson 1988). In the following, we briefly 
outline this argument, and update the derivation. 

The radius of a giant branch star depends sensitively on the mass of
the degenerate core, almost independent of envelope mass: \[ R/\rsun
\approx 9500 (M_{c}/\msun)^{4.8} .\] In contrast, the size of the
Roche lobe depends in part on the {\it total} mass $M$ of the star
through the mass ratio $q$: \[
R_{L}/\rsun = \frac{0.49 q^{2/3}}{0.6 q^{2/3} + \log (1 + q^{1/3})}
(a/\rsun)\] (Eggleton 1983). At the very
end of the Algol stage, the mass of the primary is nearly stripped to
its helium core and the mass of the secondary has been supplemented by
transfer from the original primary. When the secondary fills its Roche
lobe and enters into the common envelope phase, we can make use of the
fact that the binary separation was the same in both cases to derive:
\[ q_{f} \equiv M_{1,R} / M_{2,R} = \left( \frac{1 + \frac{5}{3}
q_{m}^{2/3} \log (1 + q_{m}^{-1/3})}{1 + \frac{5}{3} q_{m}^{-2/3} \log
(1 + q_{m}^{1/3})} \right)^{1/4.8} ,\] where $M_{1,R}$ and $M_{2,R}$
are the masses of the remnants of the original primary and secondary
respectively, and $q_{m} = M_{1,R} / (M_{2} + M_{1} - M_{1,R})$ is the
mass ratio just before the onset of the common envelope phase,
assuming conservative mass transfer during the Algol stage. Because
the Roche lobe of the (more massive) secondary is larger than that of
the remnant of the primary, the secondary is able to evolve slightly
farther than the primary.  Thus, the core of the secondary becomes
more massive than the core of the primary. The range allowed in
$q_{f}$ for helium double degenerates can be illustrated by two
cases. $q_{m}$ is maximized when $M_{1} \approx M_{2} = 1.0 \msun$ and
$M_{1,R} = 0.5 \msun$, which gives $q_{m} = 1/3$ and $q_{f}
= 0.90$.  $q_{m}$ is minimized when $M_{2} + M_{1} - M_{1,R} = 
2.25 \msun$ (the maximum secondary mass that would produce a helium white
dwarf), and $M_{1,R} \approx 0.25 \msun$, which gives $q_{m} = 0.11$
and $q_{f} = 0.81$.  So, our derived range ($q_{f} = 0.85 \pm 0.05$)
is slightly larger than the range derived by Tutukov \& Yungelson (1988).

In this discussion we have referred to
the initially most massive star as the primary. However, the star
actually observed to be brighter in the double degenerate systems is most often
referred to as the primary in the literature. Since the secondary
according to mass (initially) forms the second white dwarf, it will be
brighter in observed systems, and so would be called the
primary. Thus, for observed systems, the mass ratio signature of this
process would be $q^{\prime}_{f}=M_{bright}/M_{faint}= 1.17\pm0.06$
with the most recently formed white dwarf being more massive and
brighter than its companion.

There are four helium double degenerate systems that have adequately
measured mass ratios. The system L101-26 (WD 0957--666) is probably
the best measured of the four. It has a mass ratio
$q^{\prime}_{f}=1.15\pm0.10$ (Moran, Marsh, \& Bragaglia 1997),
providing evidence that it has been formed by the Algol/CE mechanism.
The mass ratio for the system WD 0136+768 is somewhat higher than the
theoretical predictions assuming conservative mass transfer. Although
a better estimate of the errors will be necessary to determine the
degree to which theory and observation are in disagreement, the final
mass ratio $q^{\prime}$ is expected to be less than about 1.11 for a double
degenerate created by the Algol/CE mechanism if mass is not conserved
during the Algol phase.

The remaining two systems have mass ratios that are the inverse of the
value expected for Algol/CE evolution.  The components of the
double-lined spectroscopic binary PG 1101+364 have approximately the
same surface temperatures and a mass ratio $q^{\prime}_{f}=0.87\pm0.03$
(Marsh 1995).  Because less massive white dwarfs are expected to cool
faster than more massive ones due to their larger surface areas, this
indicates that PG 1101+364 may not have formed by the Algol/CE
scenario (since the more massive and younger white dwarf would not
have been able to cool to the same temperature as the older, less
massive white dwarf) unless the temperature of the less massive white
dwarf was elevated during the CE phase.

The system L870-2 (WD 0135--052) also has a measured mass ratio
$q^{\prime}_{f}=0.86\pm0.02$, although some uncertainty exists as to
whether both components are helium white dwarfs (Saffer, Liebert, \&
Olszewski 1988). If it can be confirmed that the components are both
helium white dwarfs, then this binary would also appear to have the
component temperatures opposite the predictions of the Algol/CE
mechanism.

The most recent observations of L870-2 and PG 1101+364 suggest that
these two systems did not form as a result of a CE/CE or a CE/Algol
mechanism (see below), where the lower mass white dwarf is most
recently formed and therefore brighter.  The Algol/CE scenario is
untenable unless rejuvenation of the first-born white dwarf has taken
place (Iben et al. 1997).  In this case, it is unlikely that such
heating would be a consequence of the high entropy of the accreted
matter or gravitational compression of the white dwarf envelope since
energy would only be conducted into the white dwarf interior to a
depth set by the lifetime of the common envelope. This timescale ($<
10^2 - 10^3$ yrs) is orders of magnitude shorter than the cooling
timescale ($\sim 10^7$ yrs, see Mazzitelli \& D'Antona 1986) for the
brightest white dwarf (Feige 36, with $T_{eff} < 30,000$ K) in the
sample of double degenerate systems, therefore ruling out this
mechanism.  However, heating from the nuclear burning of hydrogen-rich
matter at the base of the non-degenerate envelope could take place.
Since the helium white dwarfs in the systems under consideration are
of low mass ($\lapprox 0.5 \msun$), this hypothesis would suggest that
such white dwarfs may have hydrogen-rich envelopes of $\gapprox
10^{-3} \msun$ (Benvenuto \& Althaus 1998).  Since the first-born
white dwarf may have residual hydrogen from its formation, this
hydrogen mass need not be accreted during its interaction with the
common envelope.  The white dwarf rejuvenation hypothesis
could be tested if the white dwarfs were found to exhibit
pulsations characteristic of ZZ Ceti variables.  It is generally
accepted that this phenomenon is due to the excitation of gravity
modes in the envelopes of white dwarf stars (see Winget et al. 1982).
Their detection could provide the means of probing the composition
structure of white dwarf envelopes to look for signs of rejuvenation.
Thus, this kind of independent measurement of the mass of the
hydrogen-rich layer of a white dwarf (as well as the white dwarf
itself) can provide important links to white dwarf asteroseismology
and may help illuminate important processes taking place during the
common envelope interaction.

The eclipsing double-line binary system HD 185510 is an example of a system 
which may be on the Algol/CE evolutionary path. 
This system is a long period ($P = 20.7 $~d; Frasca, Marilli, \& Catalano
1998) binary containing a K0 giant star and an underluminous B-type
star. Although there is still some uncertainty as to whether the B star
companion ($T = 31000$~K, $\log g \sim 7$; Jeffery \& Simon 1997) is a
helium white dwarf or a subdwarf B star, either possibility requires
that the progenitor must have had its evolution truncated while on its
first ascent of the giant branch. The long period of the system and
the relatively large mass of the K giant ($M \approx 2.3 \msun$) are
expected after an Algol-like phase of conservative (or nearly so)
mass transfer.

We can also consider the possibility that binary systems with larger
initial orbital separations ($a_{i} \gapprox 20 \rsun$) evolve through
two common envelope phases as the envelopes of the primary and
secondary are successively ejected. The final mass ratio of the white
dwarf remnants is much smaller in this scenario because the
binary separation decreases by a large factor after the first common
envelope stage. As a result, the secondary is not able to evolve as
far up the giant branch before filling its Roche lobe, and so will
leave the lower mass remnant. In contrast to the Algol/CE evolutionary
path, the secondary (in terms of initial mass) would also be hotter
and more luminous (the primary in terms of brightness) in the absence
of rejuvenation. However, the constraint that the secondary is able
evolve to a radius where it can undergo unstable mass transfer and
survive a common envelope phase is a serious one (see below).

To determine the viability of the CE/CE formation mechanism, we
estimate the maximum binary separation possible after the first common
envelope stage. For the binding energy of the envelope of a giant star, we use
an expression given by de Kool (1990): $$E_{bind} = \frac{G (M_{c} +
M_{e}) M_{e}}{\lambda R}$$ where $M_{c}$ and $M_{e}$ are the core and
envelope masses of the giant, $R$ is the giant's radius, and $\lambda$
is a constant with value $\sim 0.5$.  Assuming that the secondary must
be able to form a helium core of $0.25 \msun$ to be able to survive a
common envelope phase, a lower limit for the orbital separation after
the first common envelope phase can be obtained.  As a consequence,
the amount of orbital energy available for ejecting the envelope of
the primary can be determined. Using the equation $$E_{bind} =
\alpha_{CE} E_{orb},$$ we can then derive a lower limit for the core
mass of the {\it primary} star. Because the radius of the giant has a
relatively large dependence on core mass, this lower limit has a
relatively low dependence on the initial primary and secondary masses,
\ace, and $\lambda$.  From our assumptions we derive the
equation (to be solved numerically):
$$q_{f}^{4.8} = \frac{9.8}{q_{i}} \left(\frac{0.25}{\lambda\alpha_{CE}}
\right) \left(\frac{q_{m}}{q_{i}} - 1 \right) \left( 1 + 2.04 \frac{\log (1
+ q_{m}^{1/3})}{q_{m}^{2/3}} \right)$$ where $q_{m}=M_{2}/M_{1,R}$ and
$q_{i}=M_{2}/M_{1}$. With the constraints that $M_{2,R} > 0.25 \msun$
and $M_{2} < M_{1}$, we find that the parameter space can be limited
to $M_{1,R} > 0.47 \msun$, $M_{1}, M_{2} < 1.2 \msun$. If we use the
less strict condition $M_{2,R} > 0.2 \msun$, the limits are relaxed
substantially: the only useful mass constraint is that $M_{1,R} > 0.4
\msun$. However, the mass ratio is constrained to be
$q_{f}=q^{\prime}_{f}<0.53$ while the remnant of the secondary remains
hotter, beyond which $q^{\prime}_{f}>1.89$.  None of the observed
double degenerates are consistent with the mass ratio constraint,
although there are few helium double degenerate systems known and
selection biases could be important.

Ritter (1999) indicates that stable mass transfer from a giant star
can occur for mass ratios up to $q=M_{donor}/M_{accretor}=0.83$ in the
conservative case, and up to $q=1.2$ if mass can be lost from the
accretor in an isotropic wind having the accretor's specific angular
momentum.  If the latter holds true (since Algol systems are known to
transfer mass 
non-conservatively), then no systems having $M_{2,R} >
0.25 \msun$ could survive the CE/CE mechanism since $M_{1,R} \sim 0.5
\msun$ is a constraint on helium white dwarfs. Such a constraint on
$q$ would also cut into the parameter space allowed if $M_{2,R} >
0.2 \msun$, but would not entirely eliminate it. Mass loss from the
primary before it fills its Roche lobe would tend to make CE/CE less
likely by making stable mass transfer possible. The volume of
parameter space available for the CE/CE mechanism can be visualized
using Figure~\ref{parspace}.

According to these estimates, we conclude that binary scenarios
involving two common envelope phases ending with a detached binary
are possible only in systems with very rare combinations of initial
orbital parameters. Our hydrodynamical simulations indicate that
orbital shrinkage of at least a factor of 20 occurs as a result of the
common envelope evolution for even our most favorable case.  Provided
that the binary system survives the first common envelope phase, the
second stage of mass transfer involving the expansion of the secondary
component is most likely to occur on the subgiant branch or low on the
giant branch (with $M_{c} \lapprox 0.25 \msun$).  As the mass ratio of
the system is likely to be greater than 2 (since the first-born white
dwarf is less massive than $0.5 \msun$ and the secondary component is
greater than about $1 \msun$ to ensure nuclear evolutionary
expansion), mass transfer 
leads to a second common envelope phase with the merger of the system 
a likely outcome.

If the mass transfer from the secondary begins in the Hertzsprung gap
($P \sim 1$ day), the binary may escape a second common envelope
phase, but the unstable mass transfer is likely to lead to substantial
mass and angular momentum non-conservation.  This third mechanism
(CE/Algol) for the formation of double degenerates has been studied by
Sarna et al. (1996) under the assumption that the mass transfer
process is non-conservative. By also including angular momentum losses
associated with magnetic braking, Sarna et al. (1996) found that the
evolution of an Algol system following the common envelope phase
depended on the period after the common envelope phase.  For orbital periods 
less than about 1 day, the system evolved to shorter orbital periods
($\sim$ hours), whereas longer period systems evolved to periods
greater than 1 day.  This bifurcation in evolution is similar to that
discussed by Pylyser \& Savonije (1988) and Ergma, Sarna, \& Antipova
(1998) for low mass X-ray sources.  Although the bifurcation period is
a function of orbital parameters of the system, the results of these
studies, taken as an aggregate suggest that it is less than 1.5 days.
We note that for this mechanism the most recently formed white dwarf
(and hence the brighter member of the binary) is predicted to be lower
in mass than the first born white dwarf.  In
addition, since mass transfer takes place in the Hertzsprung gap or on
the subgiant branch, the mass of the secondary's remnant will be low
($\sim 0.15 - 0.2 \msun$ for the systems studied by Sarna et
al. 1996). If we assume that the remnant mass of the primary is at least $0.25 
\msun$ for the binary to have undergone common envelope evolution and 
survived, then the mass ratio is expected to be less than 0.8.  This upper
limit would decrease further
if the primary gained mass as a result of the stable mass transfer
(Shara, Prialnik, \& Kovetz 1993). 
As with the CE/CE mechanism, none of the
observed systems matches the mass and mass ratio constraints, although
again selection effects might be important.

The formation of double degenerate stars in the common envelope
scenario is also tested by the range of a factor of $\sim$ 100 in
orbital periods --- the shortest period known is 1.4 hr for L 101-26,
while the longest period (6.3 days) is for WD 1824+040. Because it is
unlikely that ionization energy can be tapped efficiently during the
common envelope phase, and because sufficient nuclear energy is not
likely to be available on the timescale over which matter is ejected
($\lapprox 10$ yrs), the conversion of orbital energy into binding
energy is almost certainly the prime factor in determining the final
periods for common envelope systems.  The main issue is whether this
range of periods can be explained by differences in the initial masses
and evolutionary states of the components in a single mechanism, or
whether different formation mechanisms must be invoked. The
theoretical evidence indicates that an Algol/CE scenario is a viable
formation mechanism for both long period and short period double
degenerates.  For this evolutionary channel, we expect a correlation
between final orbital period and mass of the double degenerates since
long period systems should have component masses close to $0.5 \msun$,
while short period systems could have a range of masses resulting from
differences in the mass of the secondary before common envelope. An
investigation of the CE/CE evolutionary scenario indicates that it is
a rare contributor to the double degenerate systems and could only be
of importance to the short period population at best.  The fact that
only one of the observed systems has a mass ratio consistent with
formation by the Algol/CE mechanism indicates that the evolutionary
theory is incomplete or other formation mechanisms can operate. The
possibility of the formation of double degenerates in a manner similar
to that suggested for the evolution of Algol systems where mass and
angular momentum are lost without invoking a common envelope phase
(see Sarna et al. 1996) may be fruitful.  Observational determinations
of the masses and mass ratios for long period ($\gapprox$ 1 day)
helium double degenerates would help in determining their origin in
terms of the CE/Algol or Algol/CE evolutionary channels more securely.

\subsubsection{Subdwarf B Stars\label{sdb}}

Systems containing subdwarf B (sdB) stars appear to share a number of
common characteristics (e.g., orbital period) with both
pre-cataclysmic variables and double degenerate stars. The current
observational evidence indicates that sdB stars are core helium
burning stars with masses of approximately $0.5 \msun$ (Saffer et
al. 1994).  This mass is sufficiently close to the core mass of
low-mass giants at helium flash that we should consider formation
mechanisms similar to those for systems involving helium degenerates.

Several studies (Allard et al. 1994; Theissen et al. 1995) have
indicated that the fraction of sdB stars in binaries is fairly large,
although there is still debate on the exact number.  The companions in
sdB binaries are inferred to be subgiants in many cases, although
there are at least two short-period binaries known to have main
sequence companions.  The two systems, HW Vir (Wood et al. 1993) and
PG 1336--018 (Kilkenny et al. 1998) are both eclipsing, have periods
(2.8 h and 2.4 h, respectively) in the period gap for cataclysmic
variables, and have mass ratios $q \sim 0.3$ if the sdB star has the
canonical mass of $0.5 \msun$. Intermediate-mass giants with masses of
$4.0\pm0.3 \msun$ also form hydrogen-exhausted cores that could become
sdB stars. While it is possible to create helium burning stars of $0.5
\msun$ by stripping the envelope from an intermediate mass giant, it
is very difficult to explain the abundance of sdB stars in the field
because of the relatively small number of stars expected from the
initial mass function (Saffer et al. 1994), and because sdB kinematics
are indicative of old, low-mass disk stars (Saffer 1991). In addition,
the ages of Galactic open (Liebert, Saffer, \& Green 1994) and
globular clusters (e.g. Moehler, Heber, \& Rupprecht 1996) containing
sdB stars rule out intermediate mass stars as their progenitors.

Low-mass stars provide a more natural way of creating helium burning
stars of $0.5 \msun$, although it is difficult to understand how a
star that evolves to the tip of the giant branch could lose nearly all
of its envelope mass, and yet still ignite helium in its degenerate
core. If too much mass is lost, the envelope of the giant will
collapse, the hydrogen burning shell will be extinguished, and the
core of the star will not be sufficiently massive to ignite helium in
a flash. On the other hand, if too little mass is lost, the giant will
evolve through the helium flash to the core helium burning phase
(where it would most likely be identified as a red clump star), and
eventually to the asymptotic giant branch to become a carbon-oxygen
white dwarf.  Mengel, Norris, \& Gross (1976) proposed that binary
mass transfer (which they assumed to be conservative) can create an sdB star,
although only for a small range of orbital separations.

The short periods of HW Vir and PG 1336--018 imply that a common
envelope phase has indeed occurred. A common envelope event may appear
to be incompatible with helium ignition since a giant immediately
contracts after the flash, but D'Cruz et al. (1996) find that the
giant's remnant can ignite helium provided that the core mass lies
within $\sim 0.018~\msun$~of the value at helium flash for normal
evolution.  The continuing gravitational contraction of the helium
core results in an eventual flash, even if the star has reached the
white dwarf cooling curve.  Upon ignition, the star evolves to the
helium burning main sequence. Although D'Cruz et al. (1996) examined
stars undergoing enhanced mass loss, we suggest that common envelope
evolution can create the same outcome with nearly complete ejection of
the envelope.  In this way, a giant could produce an sdB star provided that a
common envelope phase occurs near the red giant tip. If so, an sdB
star could be formed with masses as low as about $0.45~\msun$ for a
metal-rich composition (D'Cruz et al 1996). 

As is the case with the Mengel et al. mechanism, an important question
is whether this process could produce a significant number of sdB
systems. To try to answer this question, we have used a Monte Carlo
method using assumptions about binary mass and separation given in a
population synthesis study by Han (1998). We assumed a sdB binary
system could be formed if the primary star had a mass less than 2.25
\msun, the core mass of the primary was in a range of $0.018 \msun$
above $M_{c} = 0.45 \msun$. Using the core mass -- radius relation for
giants, the range in core masses corresponds to a 25\% increase in
radius. If the accretor can drive an isotropic stellar wind, then
unstable mass transfer can only occur for initial mass ratios $q =
M_{1} / M_{2} > 1.2$ (Ritter 1999). In that case, we find that the
birthrate is approximately 8 systems per 1000 years, which is more
than half the birthrate of helium double degenerate systems (from
values given in Han 1998). If there is not a stellar wind from the
accretor, the birthrate increases to 10 systems per 1000 years.
Selection effects would tend to make sdB systems appear to be more
populous than double degenerates due to their $\sim 10^8
\mbox{yr}^{-1}$ lifetimes. This value appears to make this mechanism
viable. Very precise mass measurements for the helium degenerates and
sdB stars in short period systems and continued searches for binaries
involving sdB stars can provide an important test of this formation
mechanism.

\acknowledgements

We would like to acknowledge helpful conversations with A. Cool,
E. Green, and R. Wade.  We would also like to thank the anonymous
referee for a number of useful comments. This work has been supported
by NSF grants AST-9415423 and AST-9727875.

\newpage

\figcaption{$\gamma_{CE}$ as a function of stellar radius for the
giants used in our common envelope simulations. The curves are
labeled by simulation number. \label{gamma}}

\figcaption{$yz$ density slices near the end of each of our
simulations. Density contours are 5 per decade. The velocity fields in
each panel have the same scale, with the maximum speed corresponding
to about 60 km s$^{-1}$. The times pictured are: a) 70 d for
simulation 1, b) 21 d for simulation 2, c) 900 d for simulation 3, d) 1053 d
for simulation 4, e) 373 d for simulation 5, and f) 16 d for simulation 6.
\label{rhoend}}

\figcaption{The bound mass change ($- M_{bound} /
\dot{M}_{bound}$) and orbital decay ($- a/ \dot{a}$) timescales as a
function of time for simulation 1.\label{times1}}

\figcaption{Tracers of gas mass for simulation 1. The curve labels are
as follows: $M_{tot}$, total mass; $M_{bound}$, total mass remaining
bound (including bound mass lost off of the grid); $M_{3*orbit}$, the
mass within a circular orbit three times the current binary
separation; and $M_{lost}$, mass lost from the grid. \label{mass1}}

\figcaption{Tracers of gas mass for simulation 2. The curve labels are
as follows: $M_{tot}$, total mass; $M_{bound}$, total mass remaining
bound (including bound mass lost off of the grid); $M_{vol}$,
mass in the original volume of the giant; $M_{3*orbit}$, the
mass within a circular orbit three times the current binary
separation; and $M_{lost,xy}, M_{lost,z}$, mass lost from the grid in the
(cylindrical) radial direction and in the z-direction. \label{mass2}}

\figcaption{Tracers of gas mass for simulation 3. The curve labels are
the same as in Figure~\ref{mass2}, with the addition of $M_{R/10}$,
the mass within a distance of 10\% of the original giant's radius of
the core. \label{mass3}}

\figcaption{A snaphot of the $yz$-plane at 1000 days into simulation
3. The solid lines enclose areas that are unstable according to the
H{\o}iland criterion. The dotted lines are contours of specific
entropy. \label{hoil}}

\figcaption{Tracers of gas mass for simulation 4. The curve labels are
the same as in Figures~\ref{mass1} and \ref{mass3}. \label{mass4}}

\figcaption{The spatial variation of the mass interior to a radius $R$
is illustrated for a 1 $\msun$ star at different evolutionary phases
at the base of the red giant branch.  The solid, dashed dot, and
dashed curves correspond to degenerate helium core masses of 0.19,
0.23, and 0.27 $\msun$ respectively.  Note that the range of radii for
which the mass radius profile is flat increases with larger core
masses.  The more rapid increase in mass with radius ($\gapprox 3
\times 10^9$ cm) for a star with a core mass of $0.19 \msun$ hinders
the successful ejection of the common envelope. \label{mrds}}

\figcaption{The bounding box for the parameter space available to
binaries that can survive two common envelope phases without
merging. The four panels are the edges of the space that is available
to real binaries that can create helium degenerates: $M_{1} = 2.25
\msun$, the maximum primary mass that has a degenerate helium core on
the giant branch; $M_{2} = 1.0 \msun$, the minimum mass for a
secondary that can evolve to the giant branch in the age of the disk of the Galaxy
without having accreted mass; $M_{1} = M_{2}$, the condition that the
first evolving star must be at least as massive as the secondary; and
$M_{1,R}$, the maximum mass for a helium degenerate core before helium
flash. The panels are configured in such a way that the paper can be
folded to create a wedge-shaped box with the axes being ($M_{1},
M_{2}, M_{1,R}$), and the letters a, b, and c showing points of
attachment. The condition that the secondary must reach a minimum core
mass $M_{2,R}$ for the binary to survive the second common envelope
phase is represented by solid lines showing cuts through the
boundaries of parameter space.  The heavy solid line assumes $M_{2,R}
= 0.25 \msun$, while the thin solid line assumes $M_{2,R} = 0.20
\msun$. Binaries with parameters which fall above the solid lines
would survive. The dashed line represents a mass ratio $q = 1$, while
the dotted line shows where the $q=1.2$ condition cuts the volume. If
$q>1.2$ is necessary for unstable mass transfer (and hence common
envelope), only binaries with parameters to the right of the dotted
line would survive. \label{parspace}}

\newpage
\begin{deluxetable}{ccccccccc}
\hspace*{-0.2in}
\tablecolumns{9}
\tablewidth{0pc}
\tablecaption{Orbital Parameters for Helium White Dwarf Binary Systems}
\tablehead{\colhead{} & \colhead{Name} & \colhead{$P$ (d)} & 
\colhead{$M_1(\msun)$} & \colhead{$M_2(\msun)$} &
\colhead{$q$\tablenotemark{a}} & \colhead{$T_{1}$ (K)} &
\colhead{$T_{2}$ (K)} & \colhead{References\tablenotemark{c}}}
\startdata
\cutinhead{Double Helium Degenerate Systems}
0135--052 & L870-2  &  1.56  & $0.47\pm0.05$\tablenotemark{b} & $0.52\pm0.05$\tablenotemark{b} & $0.86\pm0.02$ & 7500 & 6900 & 1\nl
0136+768  &         &  1.41  & 0.34 & 0.26 & 1.31 & & & 2\nl
0957--666 & L101-26 &  0.06  & $0.37\pm0.02$ & $0.32\pm0.03$ & $1.15\pm0.10$ & 27000 & & 3, 4 \nl
1101+364  &         &  0.15  &  0.27  & 0.31 & $0.87\pm0.03$ & 13600 & 13600 & 5, 6\nl
\cutinhead{Pre-Cataclysmic Variable Systems}
0308+096  &         &  0.29  & $0.39^{+0.13}_{-0.10}$ & $0.18\pm0.05$ & & 26200 & & 7\nl
0710+741  & GD448   &  0.10  & $0.44\pm0.03$ & $0.09\pm0.01$ & $0.22\pm0.03$ & 19000 & & 8\nl
1224+309  &         &  0.26  & $0.45\pm0.05$ & $0.28\pm0.05$ & $0.62\pm0.08$ & 29300 & & 9\nl
2256+249  & GD245   &  0.17  & $0.48\pm0.03$ & $0.22\pm0.02$ & $0.47\pm0.03$ & 22170 & & 10\nl
\enddata
\tablenotetext{a}{For the double degenerates, the mass ratio $q$ is defined 
to be the ratio of the mass of the brighter to the mass of the fainter 
component. For the pre-CV systems, it is defined to be the mass of the 
(less massive and fainter) main-sequence companion to that of the white dwarf.}
\tablenotetext{b}{It is still possible that L870-2 may be a double
degenerate system composed of carbon-oxygen white dwarfs.}
\tablenotetext{c}{1. Bergeron et al. 1989 2. Moran 1999 3. Moran et
al. 1997 4. Bragaglia et al. 1995 5. Marsh 1995 6. Bergeron et
al. 1992 7. Saffer et al. 1993 8. Maxted et al. 1998 9. Orosz et al. 1999
10. Schmidt et al. 1995 }
\label{obs}
\end{deluxetable}
\bigskip

\begin{deluxetable}{cccccc}
\tablecolumns{6}
\tablewidth{0pc}
\tablecaption{Initial Parameters of Common Envelope Sequences\label{initialt}}
\tablehead{\colhead{Sequence} & \colhead{$M_1(\msun)$} &
\colhead{$M_c(\msun)$} & \colhead{$M_2(\msun)$} & \colhead{$a_{i} (10^{12}$ cm)} & \colhead{$P_{i}$ (d)}}
\startdata
    1    &     2      &  0.355 &      0.2     & 4.0 & 34.1\nl
    2    &     1      &  0.28  &      0.35    & 2.0 & 15.4\nl
    3    &     1      &  0.45  &      0.35    & 22.0 & 560.7\nl
    4    &     1      &  0.45  &      0.1     & 20.0 & 989.2\nl
    5    &     2      &  0.45  &      0.35    & 16.0 & 263.6\nl
    6    &     1      &  0.28  &      0.45    & 1.6 & 12.8\nl
\enddata
\end{deluxetable}
\bigskip

\begin{deluxetable}{ccccccccc}
\tablecolumns{9}
\tablewidth{0pc}
\tablecaption{Final Parameters of Common Envelope Sequences\label{finalt}}
\tablehead{\colhead{} & \colhead{} & \colhead{} & \colhead{} & \colhead{} &
\multicolumn{2}{c}{\ace ~$=0.4$} & \multicolumn{2}{c}{\ace ~$=1.0$}\\
\colhead{Sequence} &
\colhead{$a_{f} (\rsun)$} & \colhead{$P_{f}$ (days)} & 
\colhead{\ace \tablenotemark{a}} & \colhead{$M_{bound,f}/\msun$} &
\colhead{$a_{f,c} (\rsun)\tablenotemark{b}$} &
\colhead{$P_{f,c}$ (days)}& \colhead{$a_{f,c} (\rsun)\tablenotemark{b}$} &
\colhead{$P_{f,c}$ (days)}} 
\startdata 
1 & 2.1 & 0.48 & 0.26 & 1.59 & 0.3 & 0.02 & 0.6 & 0.06 \nl
2 & 1.8 & 0.35 & 0.21 & 0.65 & 0.8 & 0.10 & 1.2 & 0.18 \nl
3 & 21.3 & 12.7 & 0.09 & 0.49 & 14.9 & 7.5 & 18.2 & 10.0 \nl
4 & 33.2 & 29.8 & 0.14 & 0.53 & 3.0 & 0.79 & 6.5 & 2.60 \nl
5 & 18.9 & 10.6 & 0.56 & 1.45 & 2.0 & 0.37 & 4.3 & 1.16 \nl
6 & 2.4 & 0.49 & 0.34 & 0.62 & 0.8 & 0.10 & 1.3 & 0.21 \nl
\enddata
\tablenotetext{a}{\ace as tabulated here is the efficiency of mass ejection 
as computed from the simulation when the run was terminated.}
\tablenotetext{b}{$a_{f,c}$ is an estimate of the final orbital
separation assuming that orbital energy is removed to unbind the
remainder of the gas with the efficiency \ace~ given.}
\end{deluxetable}
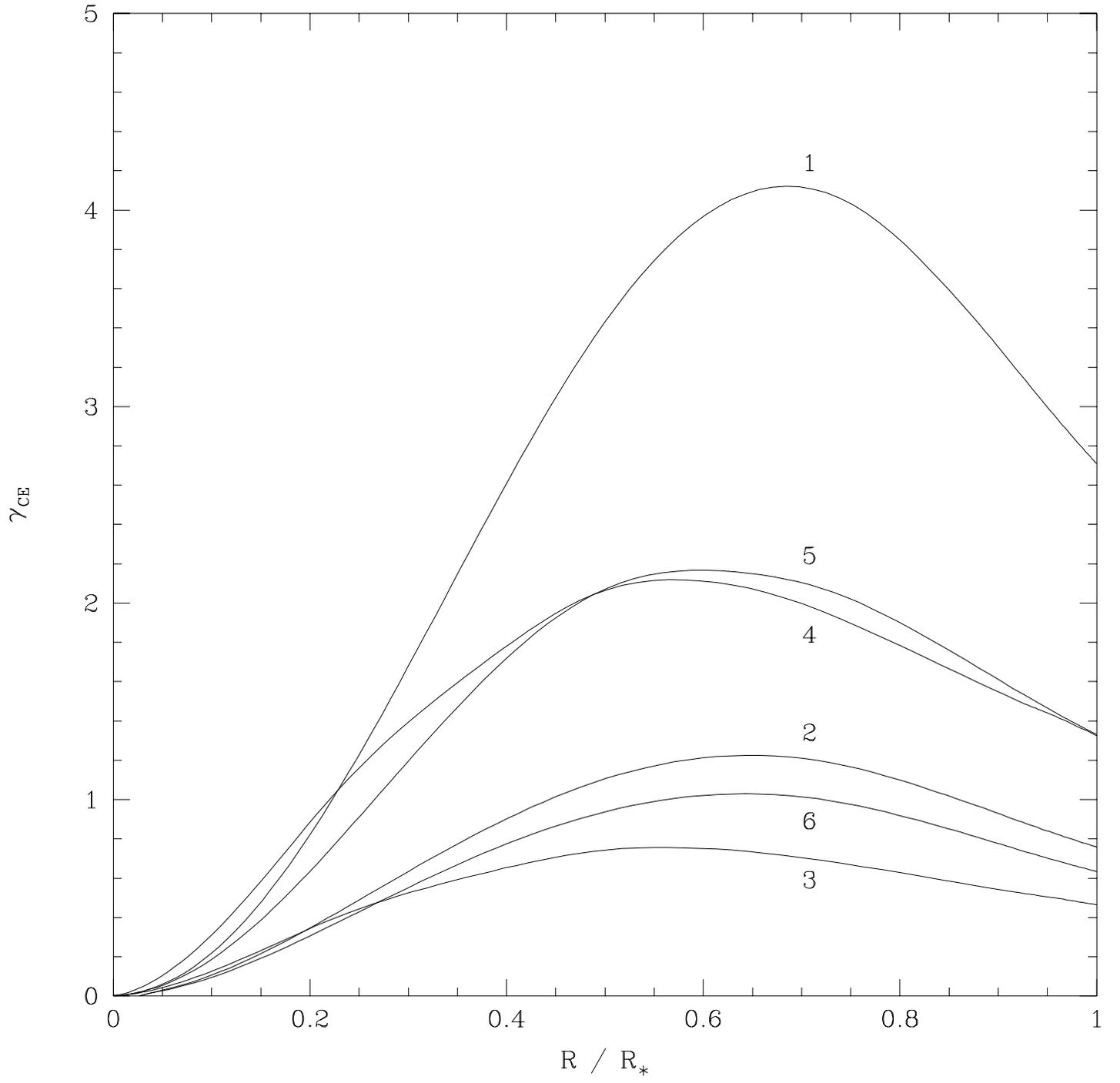
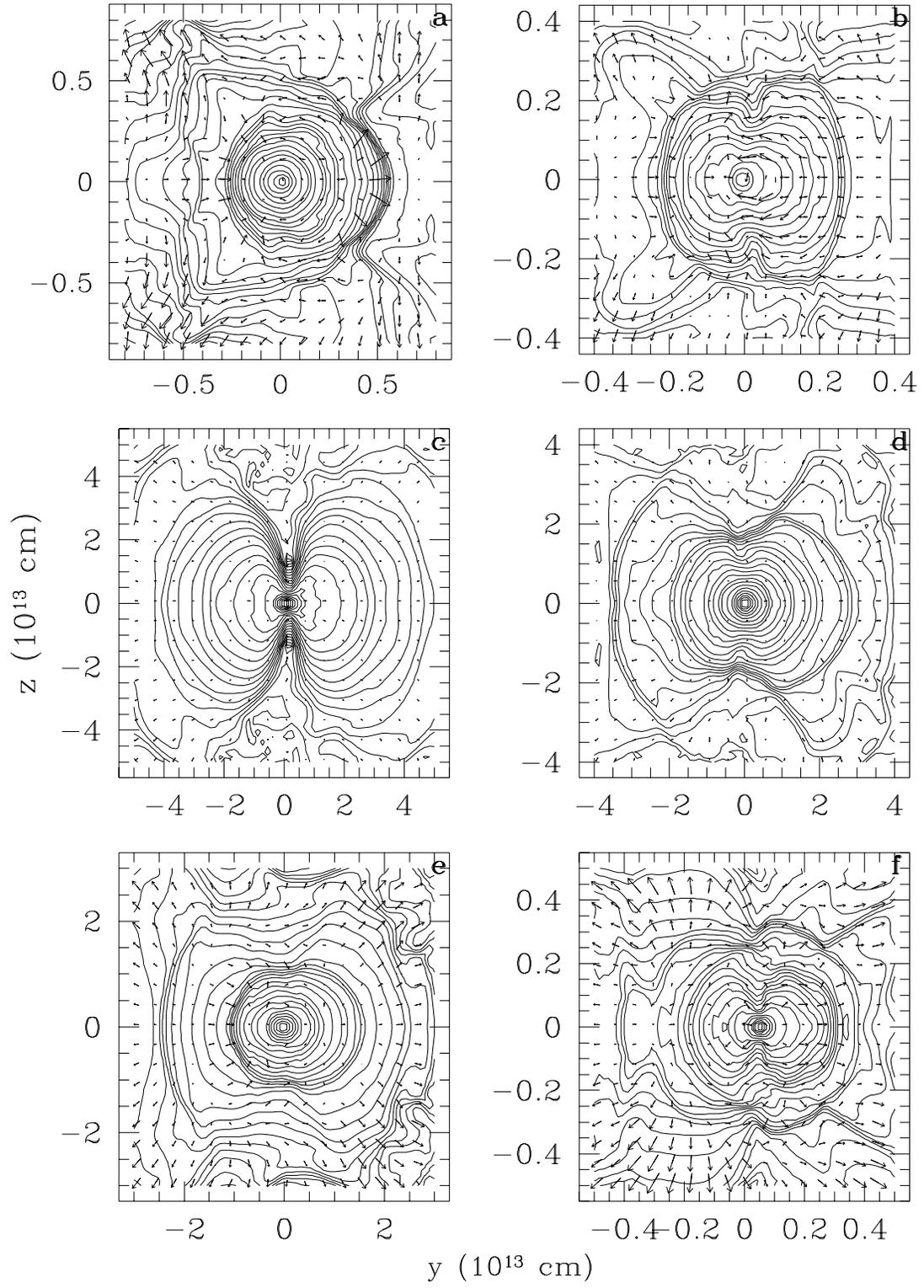
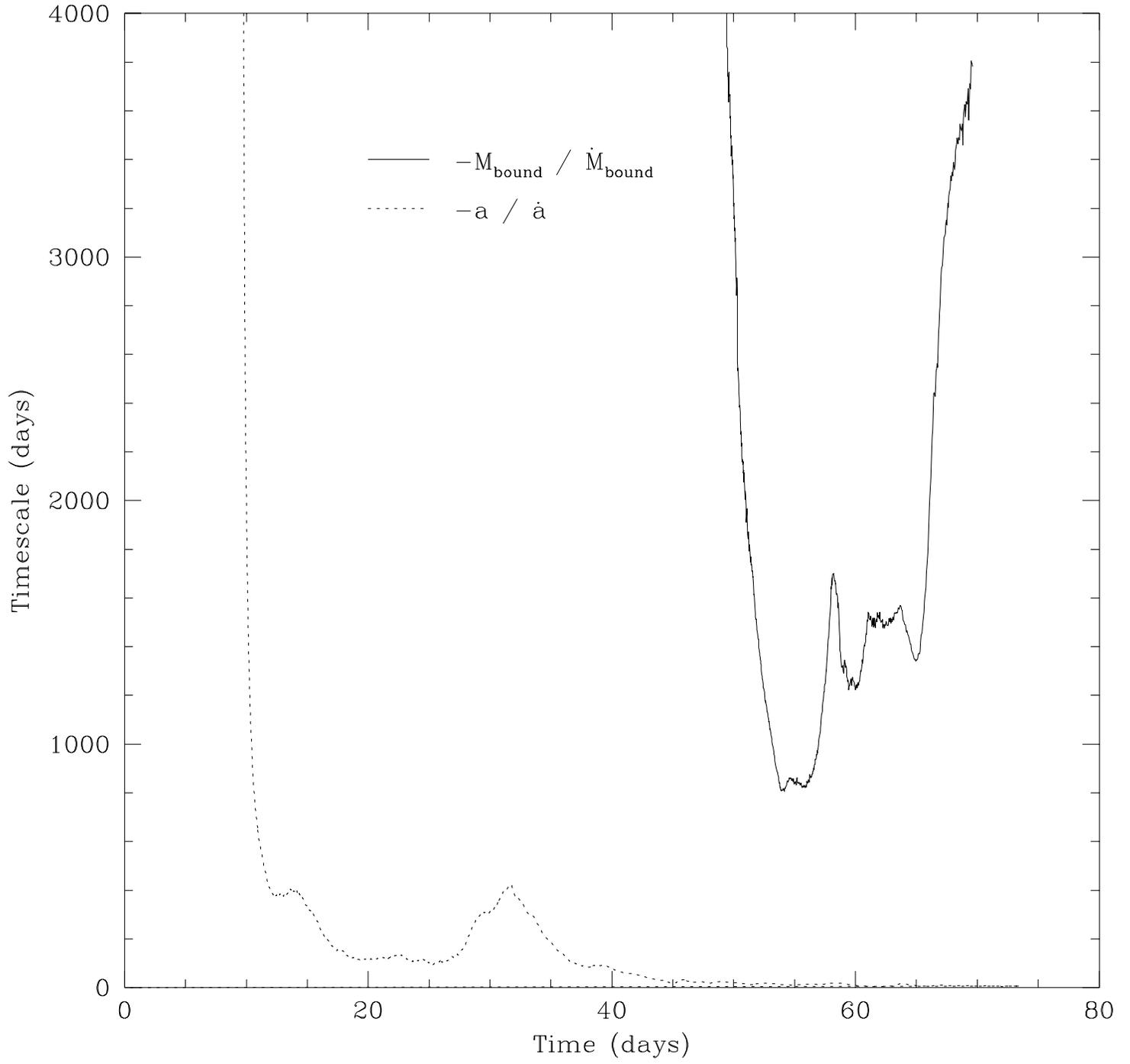
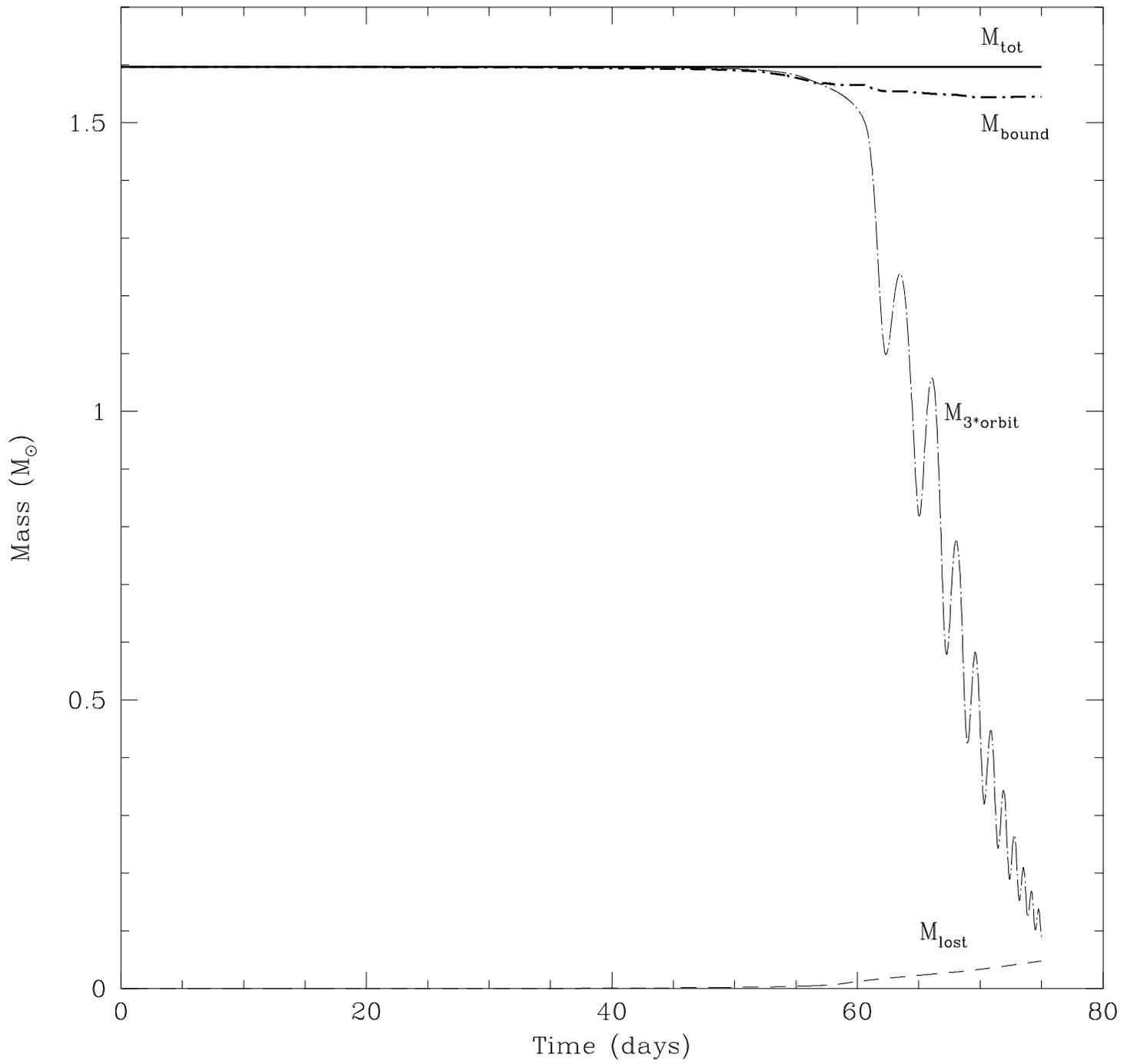
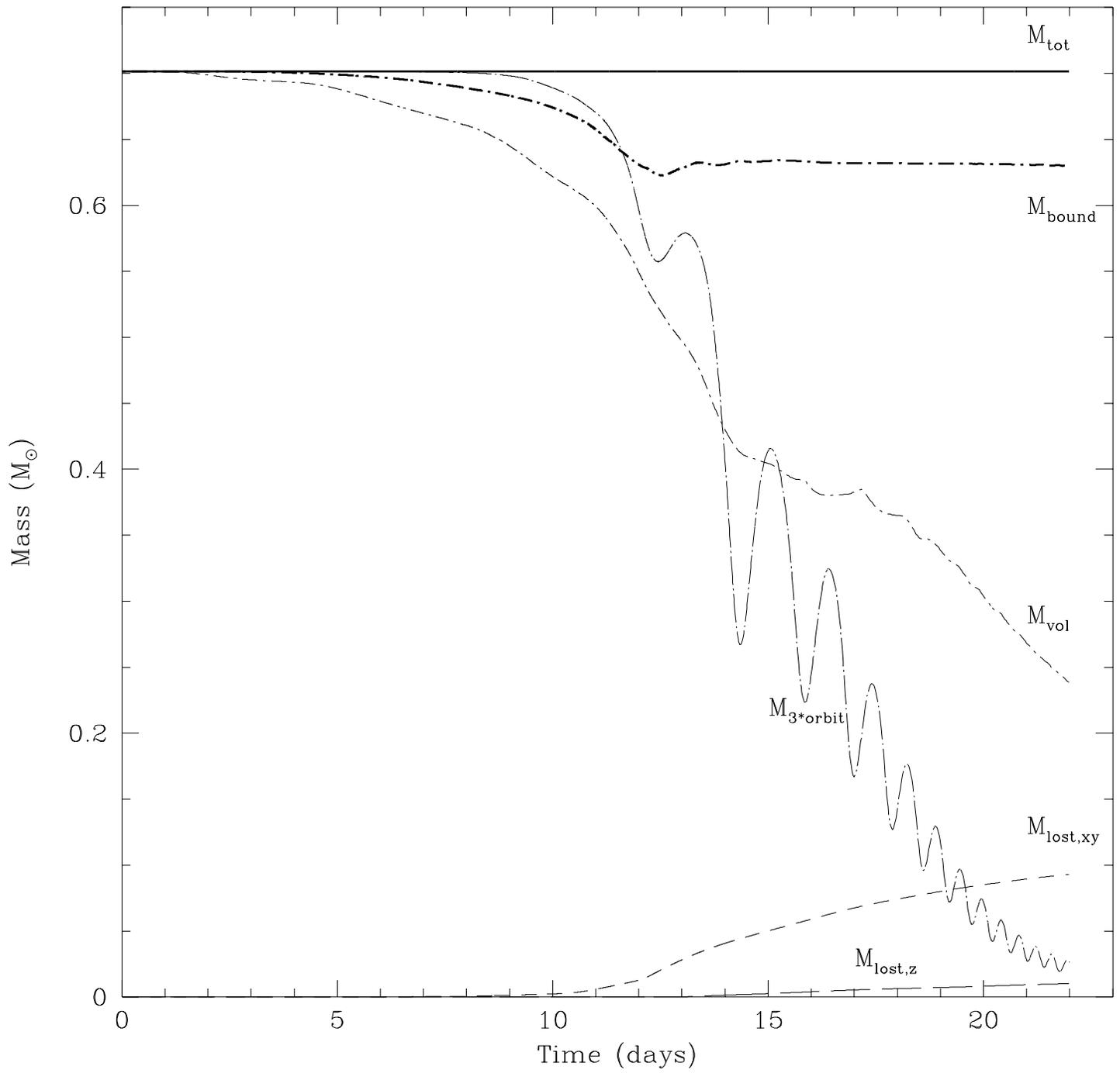
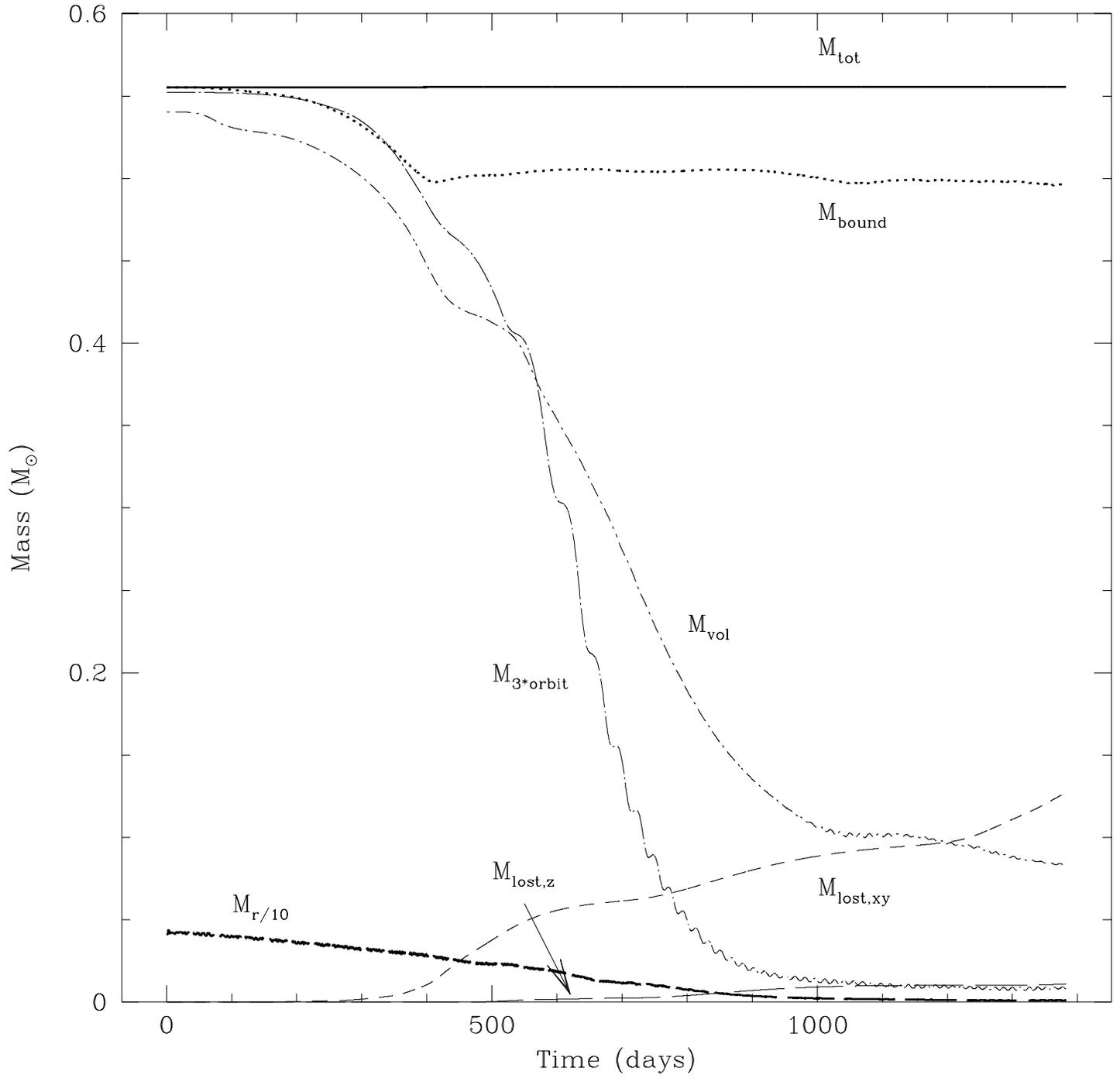
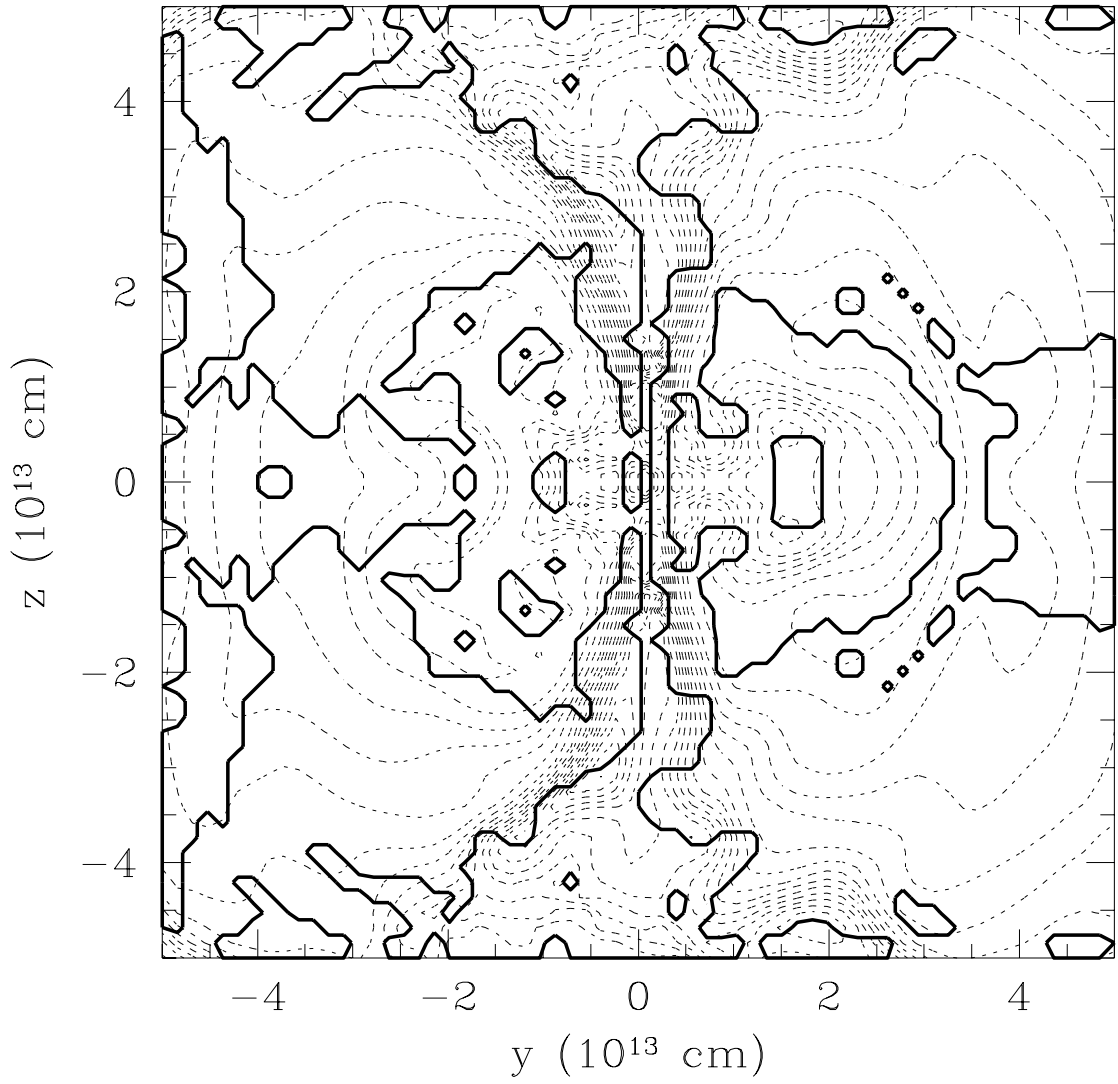
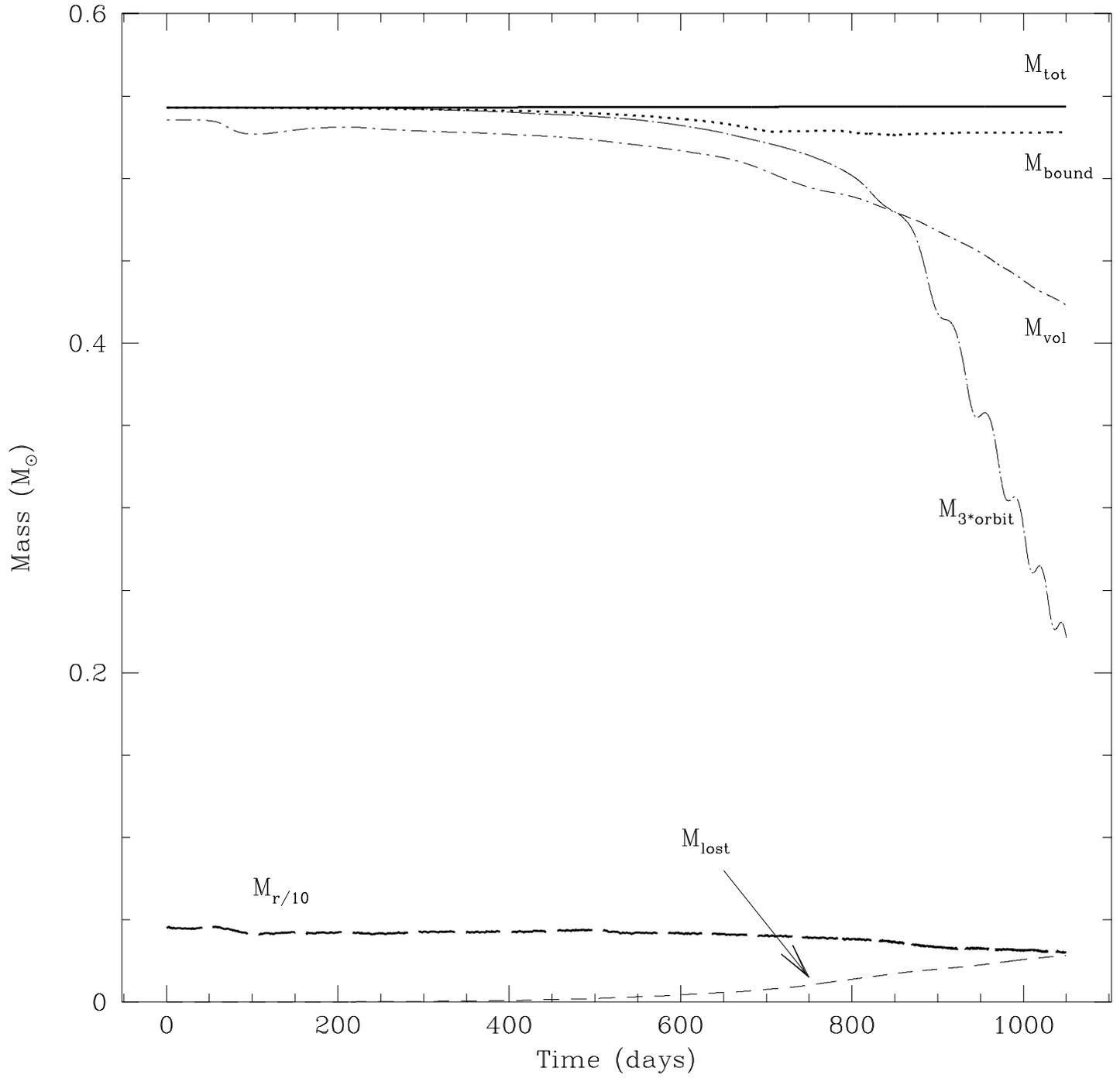
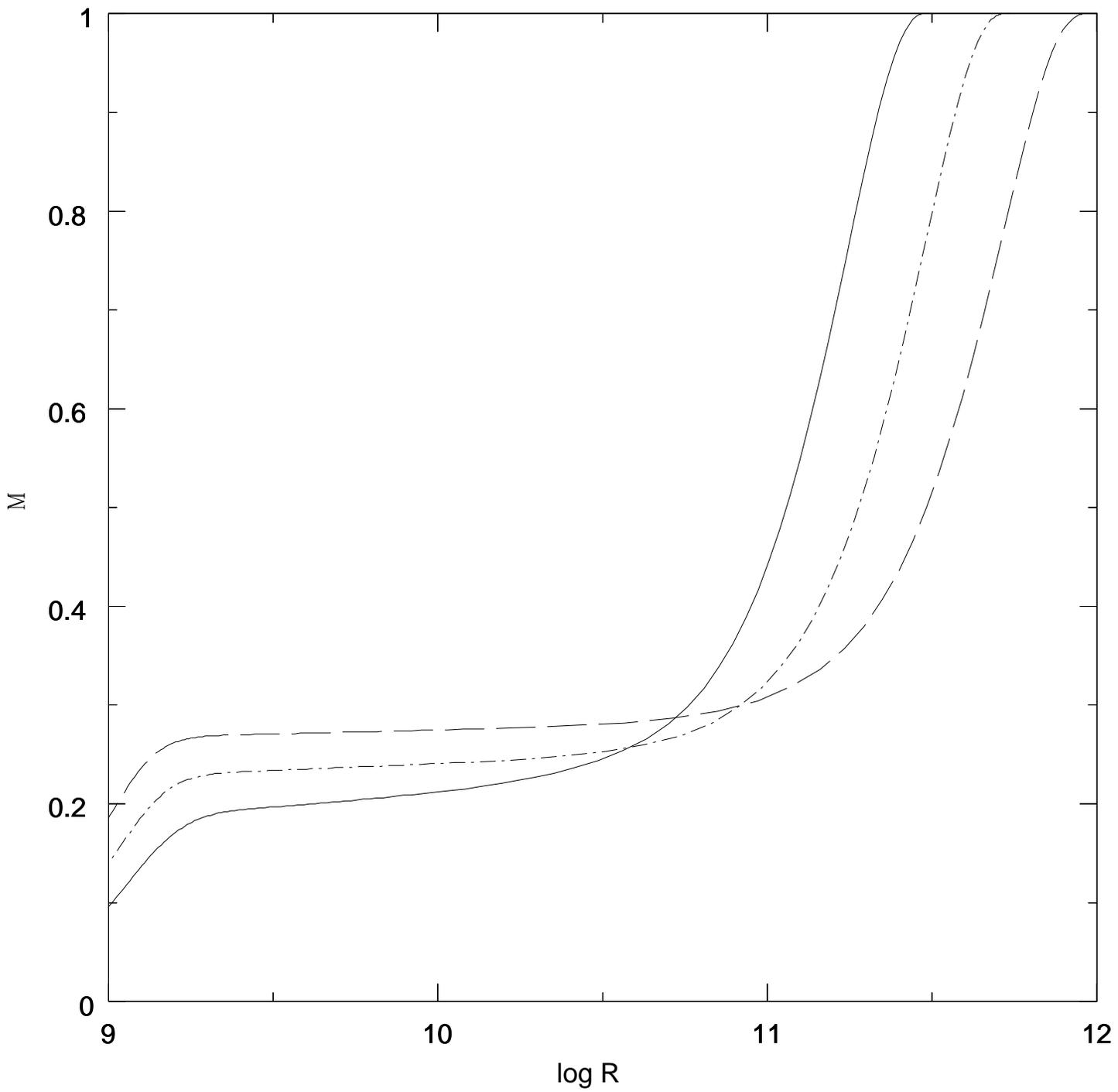
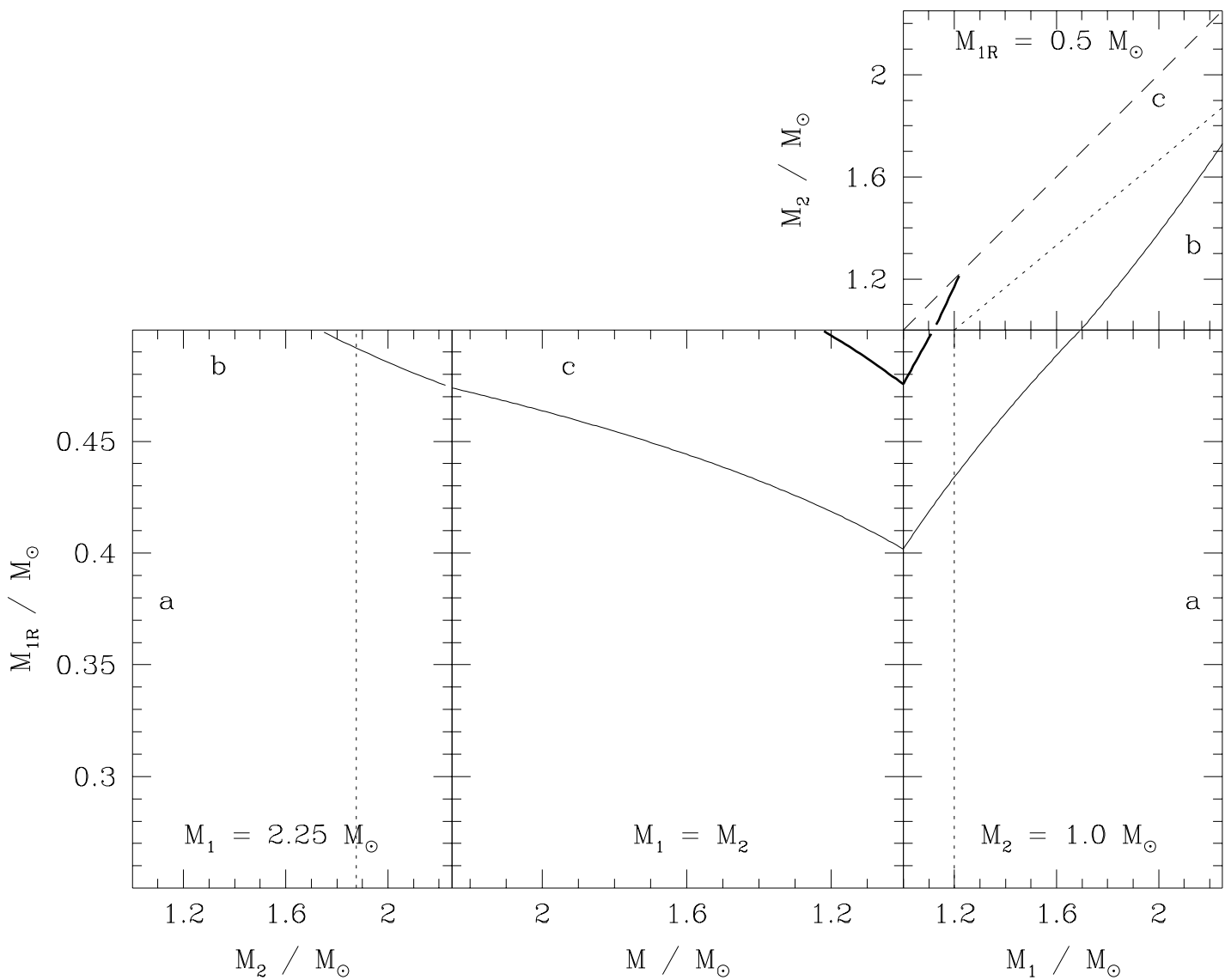
\end{document}